\documentclass[sigconf,nonacm]{acmart}
\usepackage{booktabs}\usepackage{longtable}\usepackage{array}\usepackage{multirow}\usepackage{float}
\usepackage{CJKutf8}
\begin{document}
\begin{CJK}{UTF8}{min}
\acmshorttitle{The End of Rented Discovery}
\acmshortauthors{Zhu \& Chang}
\pagestyle{acmstyle}
\thispagestyle{firstpage}

\twocolumn[
\begin{@twocolumnfalse}
\begin{center}
{\LARGE\bfseries The End of Rented Discovery: How AI Search Redistributes Power Between Hotels and Intermediaries\par}
\vspace{1.0em}

\begin{tabular}{@{}c@{\hspace{2em}}c@{}}
{\large Peiying Zhu\textsuperscript{*}} & {\large Sidi Chang\textsuperscript{*\,$\dagger$}} \tabularnewline
{\small peiying@blossomai.co} & {\small schang@blossomai.co} \tabularnewline
{\small Blossom AI} & {\small Blossom AI} \tabularnewline
{\small San Francisco, CA, USA} & {\small San Francisco, CA, USA}
\end{tabular}
\vspace{1.2em}
\end{center}

\noindent\parbox{\textwidth}{
\small
\noindent\textbf{Abstract}\\[3pt]
When a traveler asks an AI search engine to recommend a hotel, which sources get cited---and does query framing matter? We audit 1,357 grounding citations from Google Gemini across 156 hotel queries in Tokyo and document a systematic pattern we call the Intent-Source Divide. Experiential queries draw 55.9\% of their citations from non-OTA sources, compared to 30.8\% for transactional queries---a 25.1 percentage-point gap ($p < 5 \times 10^{-20}$). The effect is amplified in Japanese, where experiential queries draw 62.1\% non-OTA citations compared to 50.0\% in English---consistent with a more diverse Japanese non-OTA content ecosystem. For an industry in which hotels have long paid OTAs for demand acquisition, this pattern matters because it suggests that AI search may make hotel discovery less exclusively controlled by commission-based intermediaries.
}

\vspace{1.5em}
\end{@twocolumnfalse}
]

\renewcommand{\thefootnote}{\fnsymbol{footnote}}
\footnotetext[1]{Both authors contributed equally to this research.}
\footnotetext[2]{Corresponding author.}
\renewcommand{\thefootnote}{\arabic{footnote}}

\section{Introduction}

The hospitality distribution landscape has long been organized around a simple bargain: Online Travel Agencies help travelers discover hotels, and hotels pay for that discovery through commission. Commission rates commonly range from 15\% to 25\% of booking value \cite{dedge2024}, and for independent properties OTA-originated bookings can represent over 61\% of room revenue \cite{cloudbeds2025}. This arrangement is costly, but hotels have historically accepted it because OTAs control the comparison interface that travelers use at the moment of discovery. The economics are well documented: OTAs boost visibility and occupancy but extract high margins \cite{anderson2009billboard,ghose2012ranking}, and market entry dynamics show that brand loyalty consolidated through OTA platforms can deter independent hotel entries \cite{bianco2025deter}.

Generative AI search may change that arrangement. When a traveler asks Gemini or another AI system for a hotel recommendation, the system no longer merely ranks links; it synthesizes an answer and cites the sources it relies on. The strategic question for hotels is therefore no longer only whether they appear on OTA platforms, but whether AI systems continue to route discovery through those platforms or begin citing hotel-owned content and other non-OTA sources directly. In that sense, AI search is not simply a new marketing channel---it is a possible redistribution of who controls hotel discovery.

\subsection{GEO and Citation Audits: What We Know About AI Search}

Generative AI search engines synthesize answers from multiple sources and cite those sources as grounding references, creating a discovery architecture where visibility may depend on content relevance rather than page authority alone. The emerging GEO literature has begun to characterize these patterns: traditional SEO signals do not straightforwardly predict AI citation \cite{aggarwal2024geo}, AI search shows systematic bias toward editorial content over brand-owned content \cite{chenm2025geo}, query intent significantly shapes source selection \cite{chenx2025role}, and AI citations concentrate heavily among a small number of outlets \cite{yang2025news}. Citation audits across platforms have identified sentiment, commercial, and geographic biases \cite{li2024generative} and a ``discovery gap'' in which discovery-style queries produce very different citation patterns than named-entity queries \cite{sharma2026discovery}. Our study adapts this citation-audit methodology for the hospitality domain.

\subsection{Theoretical Grounding: Query Intent and Source Diversity}

Our study draws on two established lines of research. First, the information retrieval literature has long recognized that query intent shapes search behavior and result characteristics. Broder~\cite{broder2002taxonomy} proposed a foundational taxonomy distinguishing navigational, informational, and transactional queries, demonstrating that different intent types activate different retrieval strategies and satisfy different information needs. Our transactional/experiential distinction maps onto this taxonomy: transactional hotel queries resemble Broder's transactional class (goal: complete a specific action), while experiential queries resemble the informational class (goal: acquire knowledge about a topic). The key insight from this literature is that query intent is not merely a user characteristic but a structural feature of the query that determines what kinds of web content constitute a satisfactory answer.

Second, the algorithm auditing literature \cite{introna2000shaping,kulshrestha2017quantifying} has established that search engine ranking algorithms embed systematic biases in which information sources gain visibility. Our study extends this tradition to generative AI search, where the mechanism shifts from ranking to citation---the AI system does not rank sources but selects which to cite as grounding for its synthesized response. The citation-audit methodology we employ is conceptually equivalent to the ranking-audit methodology of traditional search engine studies, adapted for the generative paradigm.

These two strands converge in a simple theoretical prediction: if AI search operates as a content-matching system (retrieving sources whose content best matches the query's information need), then queries with different intent types should systematically retrieve different source types, because the web content that best answers a transactional query (structured data, price comparisons, booking interfaces) differs from the content that best answers an experiential query (narrative descriptions, editorial recommendations, atmosphere characterizations). The intent-source divide, if observed, would be consistent with content-matching playing a significant role in AI search citation, though it would not rule out authority signals as a contributing factor.

\subsection{The Gap: No Study Has Done This for Hospitality}

Despite growing work on AI search citation patterns and on hotel-OTA distribution economics, to our knowledge, no study has examined how AI search changes intermediation in hospitality. This gap matters because hotels are a rare setting where the discovery intermediary is clearly identifiable and economically important. If Gemini cites Booking.com, Expedia, or Jalan, the hotel is being surfaced through a commission-based intermediary. If Gemini cites the hotel's own website, the hotel is competing for discovery more directly. Citation source does not tell us exactly where the guest will book, but it does tell us who captured the discovery moment at the moment the recommendation was made. In a market where discovery has long been rented from OTAs, that distinction has direct strategic and economic significance. Additionally, hotels have a spatial dimension that interacts with query intent (a station-access query differs from a neighborhood-atmosphere query), and the three-tier market structure (international chains, domestic chains, independents) creates natural variation in web presence, content depth, and OTA dependency.

\subsection{Why Tokyo}

We study this question in Tokyo for three reasons that make it an ideal natural laboratory. First, Tokyo's hotel market exhibits a well-defined three-tier structure: international chains (Marriott, Hyatt, Hilton, InterContinental) with global brand websites optimized for English-language discovery; domestic chains (APA Hotels, Toyoko Inn, Dormy Inn, Mitsui Garden Hotels) with content-rich Japanese-language websites serving a large domestic market; and independent properties (ryokans, boutique hotels, capsule hotels, design hotels) with highly variable web presences. This three-tier structure generates natural variation in website content depth, language coverage, and OTA dependency that allows us to examine whether AI citation patterns systematically differ across market segments.

Second, Tokyo operates within two linguistically separate web ecosystems. English-language and Japanese-language web content about Tokyo hotels are largely distinct corpora: they feature different domains, different editorial voices, different OTA platforms (Booking.com and Expedia dominate the English web; Jalan, Rakuten Travel, and Ikyu dominate the Japanese web), and different non-OTA content types. Because Google Search grounding---the mechanism through which Gemini retrieves source material---searches the web in the query language, English and Japanese queries effectively surface different information markets. This bilingual structure allows us to test whether AI citation patterns reflect properties of the underlying web ecosystem rather than properties of the AI model itself.

Third, Tokyo is one of the world's largest inbound tourism markets: Japan recorded 36.9 million foreign visitors in 2024, the highest figure in the country's history, with Tokyo as the primary gateway \cite{jnto2025}. Hotel discoverability in Tokyo has substantial economic consequences. The combination of high visitor volume, a competitive hotel market, and the coexistence of international and domestic booking platforms makes Tokyo a setting where the question of AI search intermediation is both empirically tractable and practically significant.

\subsection{Research Questions}

We pose three research questions: (1)~\textbf{Discovery intermediation by intent.} Does query framing shift AI hotel discovery away from OTA intermediaries? (2)~\textbf{Language and ecosystem structure.} Does the shift depend on the language-specific web ecosystem? (3)~\textbf{Hotel-direct citation and content depth.} Do hotel-owned websites appear in AI citations, and what content characteristics appear to distinguish cited from non-cited properties?

These questions are addressed through a systematic audit of 1,357 grounding citations from 156 queries to Gemini 2.5 Flash, using a paired query design that isolates the effect of intent framing on source selection. This is a hospitality strategy paper about AI-era hotel discovery. It uses GEO concepts and citation-audit methods as the mechanism and methodology, but its core contribution is to the question of who controls hotel discovery as the search interface shifts from link-based ranking to AI-synthesized recommendation.

\section{Data and Methods}

\subsection{Query Design}

We designed a structured query corpus of 156 queries organized as category-matched pairs. Each pair consists of a transactional variant and an experiential variant addressing the same traveler need category in the same geographic area, executed in both English and Japanese. This paired design ensures that the only variable changing between queries within a pair is the intent framing---not the topic, geographic scope, or language---allowing us to isolate the effect of intent on citation source composition.

\textbf{Query categories.} We identified four traveler need categories that span the primary dimensions of hotel search: (1) \textit{Budget}---price-sensitive discovery, where OTAs have a strong structural advantage due to price comparison functionality; (2) \textit{Rating/Quality}---quality-oriented discovery, where aggregated review scores and star ratings are the primary signals; (3) \textit{Convenience}---location and access-oriented discovery, where proximity to transit nodes and walkability are the key criteria; and (4) \textit{Business}---work-oriented discovery, where workspace quality, quiet environments, and business amenities define the need. These categories were selected to cover the range of functional and experiential dimensions that prior research on hotel choice behavior has identified as primary decision factors.

For each category, the transactional variant was designed to resemble a booking-oriented query---the kind of query that would naturally lead to an OTA listing (e.g., ``Cheap hotel in Shinjuku,'' ``Best rated hotel in Shibuya''). The experiential variant was designed to describe the same need in terms of the guest experience rather than a bookable attribute (e.g., ``Good value hotel with local charm in Shinjuku,'' ``Hotel with exceptional service and atmosphere in Shibuya''). Japanese translations were crafted to be natural-sounding queries rather than literal translations, preserving the intent distinction in idiomatic Japanese (e.g., transactional: ``新宿で安いホテル''; experiential: ``新宿で地元の雰囲気が楽しめるコスパの良いホテル'').

\textbf{Geographic scope.} Area-level queries were executed across 9 of Tokyo's 23 special wards, selected on the basis of hotel density and tourism significance. The nine wards comprise the Toshin 5-ku (都心5区) core---Chiyoda, Chuo, Minato, Shinjuku, and Shibuya---which constitute Tokyo's central business and tourism district, plus four additional wards with high concentrations of tourist-relevant hotels: Taito (Asakusa and Ueno, the traditional ryokan district), Toshima (Ikebukuro, a major budget hotel hub), Shinagawa (a Shinkansen gateway and business hotel cluster), and Koto (Toyosu and Odaiba, with newer hotel developments). The remaining 14 wards were excluded because they contain relatively few tourist-oriented hotels and would have generated queries about predominantly residential areas, introducing noise without proportionate analytical value. In addition to the 144 area-level queries (4 categories $\times$ 2 intents $\times$ 9 wards $\times$ 2 languages), we included 12 city-level queries (3 categories $\times$ 2 intents $\times$ 2 languages) that address Tokyo as a whole, yielding a total of 156 queries. Appendix~B provides sample queries illustrating the paired design.

\subsection{Data Collection}

All 156 queries were executed on Gemini 2.5 Flash with Google Search grounding enabled in March 2026, using default generation parameters (temperature = 1.0, no seed; see Appendix~A for full configuration). When grounding is active, Gemini issues Google Search queries in the input language, retrieves top results, and synthesizes a response citing the consulted web pages. Each query was executed once; because Gemini's grounding output is non-deterministic, we additionally re-ran a stratified subset of 20 queries five times each to assess test-retest reliability (see Appendix~E). We extracted 1,357 grounding citations across 156 queries, yielding a mean of 8.7 citations per query (range: 2--32, CV: 53\%). To address variability in citation counts, we employ both citation-weighted and query-weighted analytical approaches (see Section~2.5).

\subsection{Source Classification}

Each grounding citation was classified as either OTA (booking-oriented intermediary, including global, Japanese, APAC, and meta-search platforms) or non-OTA. Non-OTA sources were further classified into nine sub-types: hotel direct, editorial curation, travel blog, travel media, local tourism, coworking/workspace, travel agency, user-generated content, and accommodation platform. Classification was by domain matching against curated lists (see Appendix~C).

\subsection{Hotel Name Extraction and Tier Classification}

Hotel names were extracted from Gemini's responses and classified into three tiers: \textbf{international chain} (Marriott, Hyatt, Hilton, IHG, Accor and sub-brands), \textbf{domestic chain} (APA Hotel, Toyoko Inn, Dormy Inn, Super Hotel, Mitsui Garden, Prince Hotels, and others), or \textbf{independent}. Classification follows the guest-facing brand name---the name a traveler would recognize---not the parent company. A hotel qualifies as ``chain'' if its brand name appears on three or more properties---a threshold that captures the brand recognition and cross-property consistency relevant to traveler search behavior, while excluding one-off coincidental name overlaps.

\subsection{Statistical Methods}

We test the Intent-Source Divide using chi-squared tests, odds ratios with 95\% confidence intervals, and Cramer's V for effect size. To adjust for language and query category, we estimate logistic regression models at the citation level ($n = 1{,}357$) and a query-level quasibinomial specification ($n = 156$) to address non-independence of citations within queries. A Mann-Whitney U test provides a query-weighted robustness check. Full test details, coefficients, and robustness results are reported in Appendices~D--E.

Citations are clustered within queries and shared templates, so citation-level p-values and confidence intervals should be interpreted with caution. The query-level model confirms that results are substantively unchanged when this clustering is accounted for.

\section{Results}

Across 156 Gemini 2.5 Flash queries, we extracted 1,357 grounding citations. OTAs account for 55.3\% of all citations (751 of 1,357), with the remainder distributed across hotel direct websites, editorial platforms, travel blogs, and other non-OTA sources. We observed systematic variation in this OTA/non-OTA split by query intent, language, and category. Because citations are partially clustered within queries and shared templates, the citation-level results below should be interpreted as substantively informative but somewhat optimistic in precision. The main pattern, however, is large in magnitude and is corroborated by query-level robustness checks.

\subsection{The Intent-Source Divide}

The central empirical finding is a substantial difference in non-OTA citation rates between experiential and transactional hotel queries. Non-OTA sources account for 55.9\% of experiential citations (419 of 750) but only 30.8\% of transactional citations (187 of 607), a gap of 25.1 percentage points. A chi-squared test confirms that this difference is highly significant: $\chi^2(1) = 84.23$, $p = 4.40 \times 10^{-20}$. The unadjusted odds ratio is 2.84 (95\% CI [2.27, 3.56]): experiential queries are nearly three times as likely to cite non-OTA sources as transactional queries. The association is moderate in size by Cramer's V ($V = 0.249$).

To examine whether this pattern holds after accounting for language and query category, we estimated two logistic regression models predicting the probability that a citation comes from a non-OTA source. Full coefficients are reported in Appendix~D (Tables~\ref{tab:logistic}--\ref{tab:quasibinomial}). In the main-effects model (Model~1), experiential intent is the strongest predictor (adjusted OR = 2.95, 95\% CI [2.34, 3.71], $p < 0.001$). Japanese queries also produce significantly more non-OTA citations than English queries (OR = 1.33, 95\% CI [1.06, 1.67], $p = 0.015$). Relative to budget queries, business and convenience queries produce significantly more non-OTA citations (business OR = 2.57, convenience OR = 2.00).

Model~2 adds an interaction between experiential intent and Japanese-language queries. This interaction is significant (OR = 1.77, 95\% CI [1.12, 2.80], $p = 0.015$; AIC 1732.1 vs.\ 1736.0; likelihood-ratio $\chi^2(1) = 5.90$, $p = 0.015$), indicating that the non-OTA boost from experiential framing is amplified in Japanese. For Japanese-language experiential queries, 62.1\% of citations come from non-OTA sources---nearly double the 31.8\% observed for English transactional queries (direct cell comparison: OR = 3.52, 95\% CI [2.53, 4.90], $p < 10^{-14}$). This is consistent with the descriptive language split reported in Section~3.2.

The intent-source divide is robust across all four query categories. Table~\ref{tab:category_intent} presents the category-level breakdown:

\begin{table*}[t]
\caption{Non-OTA Citation Rate by Query Category and Intent}
\label{tab:category_intent}
\centering
\small
\begin{tabular}{lrrrcrl}
\toprule
Category & T non-OTA\% & E non-OTA\% & Gap (pp) & 95\% CI & $\chi^2(1)$ & $p$-value \\
\midrule
Budget & 15.5\% & 42.5\% & 27.0 & [17.5, 36.5] & 26.00 & $3.4 \times 10^{-7}$ \\
Rating/Quality & 22.1\% & 46.7\% & 24.6 & [13.6, 35.5] & 16.78 & $4.2 \times 10^{-5}$ \\
Business & 40.1\% & 68.5\% & 28.4 & [18.1, 38.7] & 25.81 & $3.8 \times 10^{-7}$ \\
Convenience & 40.1\% & 59.2\% & 19.0 & [9.3, 28.8] & 13.29 & $2.7 \times 10^{-4}$ \\
\bottomrule
\end{tabular}
\end{table*}

Several patterns are notable. Budget queries show the lowest non-OTA share in the transactional condition (15.5\%), consistent with the structural advantage that OTA platforms hold in price-comparison functionality. Business queries show the largest absolute gap (28.4 percentage points) and the highest non-OTA share in the experiential condition (68.5\%), driven by coworking review platforms, hotel direct websites describing workspace amenities, and specialized business travel content that OTA listings do not provide. Convenience queries show the smallest gap (19.0 pp), suggesting that location and access information is somewhat available across both OTA and non-OTA sources. All four category-level gaps remain significant under Bonferroni correction (adjusted $\alpha = 0.0125$).

The intent-source divide is robust to query-weighted analysis, in which each query contributes equally regardless of citation count (query-weighted non-OTA rates: 55.1\% experiential vs.\ 27.1\% transactional; Mann-Whitney $U = 4{,}721$, $p < 10^{-9}$). It also survives controls for query length and lexical richness, answer-type specificity tests, and test-retest replication across five independent runs (ICC = 0.656). Full robustness results are reported in Appendix~E; a summary of all statistical tests is in Appendix~D (Table~\ref{tab:test_summary}).

\subsection{Language, Ecosystem Structure, and Non-OTA Composition}

The intent-source divide operates differently across languages. Table~\ref{tab:language_intent} presents the four-segment breakdown:

\begin{table}[t]
\caption{OTA and Non-OTA Citation Rates by Language and Intent}
\label{tab:language_intent}
\begin{tabular}{lrrr}
\toprule
Segment & OTA\% & Non-OTA\% & $N$ \\
\midrule
EN Transactional & 68.2\% & 31.8\% & 274 \\
EN Experiential & 50.0\% & 50.0\% & 386 \\
JP Transactional & 70.0\% & 30.0\% & 333 \\
JP Experiential & 37.9\% & 62.1\% & 364 \\
\bottomrule
\end{tabular}
\end{table}

The most striking cell is Japanese experiential: 62.1\% of citations come from non-OTA sources. In English, the intent gap is 18.2 percentage points; in Japanese, 32.1 percentage points---nearly twice as large. The interaction term confirms this amplification (OR = 1.77, $p = 0.015$). The divergence is concentrated in experiential queries (EN 50.0\% OTA vs.\ JP 37.9\%, a 12.1 pp gap) rather than transactional queries (EN 68.2\% vs.\ JP 70.0\%, a 1.8 pp gap).

This amplification is consistent with the structure of two effectively separate web ecosystems. Japanese queries cite Japanese-domain pages 68.4\% of the time; English queries cite Japanese-domain pages only 6.4\%. The non-OTA channel accounts for 42.4\% of English citations (280 of 660) and 46.8\% of Japanese citations (326 of 697), with significantly different composition ($\chi^2(10) = 171.90$, $p < 0.001$). Table~\ref{tab:nonota_composition} presents the breakdown:

\begin{table}[t]
\caption{Non-OTA Source-Type Composition by Language (\% of Non-OTA Citations)}
\label{tab:nonota_composition}
\begin{tabular}{lrr}
\toprule
Source Type & EN & JP \\
\midrule
Travel blog & 22.9\% & 1.2\% \\
Editorial curation & 22.5\% & 12.3\% \\
Hotel direct & 19.3\% & 23.6\% \\
Travel agency & 0.0\% & 12.9\% \\
Travel media & 8.9\% & 3.7\% \\
UGC & 6.4\% & 8.6\% \\
Coworking/workspace & 2.5\% & 8.3\% \\
Local tourism & 5.4\% & 7.7\% \\
Accommodation platform & 2.5\% & 4.3\% \\
Other & 8.9\% & 17.5\% \\
\bottomrule
\end{tabular}
\end{table}

The English non-OTA channel is dominated by travel blogs (22.9\%) and editorial curation sites (22.5\%)---international content creators writing about Tokyo for foreign tourists. The Japanese non-OTA channel is more diverse: hotel direct websites are the largest category (23.6\%), but the most distinctive feature is the presence of source types absent from English---travel agency sites (12.9\% vs.\ 0.0\%) and coworking/workspace platforms (8.3\% vs.\ 2.5\%). These are not better or worse ecosystems---they are differently structured, with different source types available to absorb the OTA share when queries become experiential.

\subsection{Hotel Direct Citation: Language Matters More Than Intent}

A more nuanced pattern emerges for hotel-direct citation. Within each language, we do not detect a clear intent effect: in English, hotel-direct citation rates are nearly identical for transactional and experiential queries (8.0\% vs.\ 8.3\%, $p = 0.90$), while in Japanese the experiential rate is higher than the transactional rate (12.6\% vs.\ 9.3\%), though this within-language difference is not statistically significant ($p = 0.16$). Table~\ref{tab:hotel_direct} presents hotel-direct citations as a share of all citations by segment.

\begin{table*}[t]
\caption{Hotel Direct Citation Rate (\% of All Citations) by Language and Intent}
\label{tab:hotel_direct}
\centering
\small
\begin{tabular}{lrrrrl}
\toprule
Language & T-rate & E-rate & Difference & 95\% CI & $p$ \\
\midrule
English & 8.0\% ($n$=274) & 8.3\% ($n$=386) & +0.3 pp & [$-$4.0, 4.5] & 0.90 \\
Japanese & 9.3\% ($n$=333) & 12.6\% ($n$=364) & +3.3 pp & [$-$1.3, 8.0] & 0.16 \\
\bottomrule
\end{tabular}
\end{table*}

The stronger pattern is cross-linguistic. Aggregating across intent, hotel-direct websites account for 8.2\% of all English citations and 11.0\% of all Japanese citations. Although this 2.8 percentage-point gap falls short of conventional significance in a citation-level test ($\chi^2(1) = 2.87$, $p = 0.090$), the query-level model---which avoids pseudo-replication across citations within the same response---shows that Japanese queries are significantly more likely than English queries to produce any hotel-direct citation (OR = 2.27, 95\% CI [1.08, 4.75], $p = 0.030$), controlling for intent and category.

This cross-linguistic difference is especially visible in experiential search. Among experiential queries, hotel-direct citation is 12.6\% in Japanese compared with 8.3\% in English (one-sided z-test, $p = 0.026$). Taken together, these results suggest that hotel-direct capture is shaped less by transactional versus experiential framing alone than by the structure and depth of the underlying language ecosystem. This is consistent with the possibility that Japanese hotel websites contain richer search-answerable content, a possibility explored in the content audit below.

\subsection{Exploratory Content Audit: What Characterizes Cited Hotels?}

To explore what distinguishes cited from non-cited hotels, we audited the websites of seven hotels that Gemini cites directly and seven control hotels that were not cited. We scored each hotel on a search-answerable depth (SAD) scale across five dimensions (FAQ, area guide, blog, access information, unique/distinctive content), each scored 0--3 based on depth of content available for Google Search to retrieve: \textbf{0}~=~absent; \textbf{1}~=~present but shallow (e.g., a FAQ with $<$10 questions); \textbf{2}~=~moderate depth (e.g., 10--25 FAQ questions or area content covering several neighborhoods); \textbf{3}~=~deep (e.g., a 30+ question FAQ organized by topic, or dedicated neighborhood pages with 500+ words each). The key distinction is not whether a content feature exists, but whether it is deep enough to rank in Google Search for relevant traveler queries. Table~\ref{tab:sad_audit} presents the results.

\begin{table*}[t]
\caption{Search-Answerable Depth Audit (14 Hotels)}
\label{tab:sad_audit}
\centering
\small
\begin{tabular}{lccccccc}
\toprule
Hotel & Cited & FAQ & Area Guide & Blog & Access & Unique Content & SAD Score \\
\midrule
Kimpton Shinjuku & Yes & 3 & 3 & 2 & 2 & 3 & 13/15 \\
Kadoya Hotel & Yes & 3 & 3 & 2 & 2 & 3 & 13/15 \\
Super Hotel & Yes & 2 & 2 & 2 & 2 & 2 & 10/15 \\
Park Hotel Tokyo & Yes & 1 & 1 & 1 & 1 & 3 & 7/15 \\
Tokyu Stay & Yes & 1 & 0 & 0 & 2 & 3 & 6/15 \\
TRUNK (Hotel) & Yes & 0 & 2 & 1 & 1 & 2 & 6/15 \\
Prince Hotels & Yes & 0 & 0 & 0 & 3 & 2 & 5/15 \\
Hotel K5 & No & 1 & 1 & 1 & 1 & 1 & 5/15 \\
NOHGA Hotel Ueno & No & 0 & 2 & 0 & 1 & 2 & 5/15 \\
Ryokan Shigetsu & No & 1 & 1 & 0 & 1 & 1 & 4/15 \\
Aoyama Grand Hotel & No & 1 & 0 & 1 & 1 & 1 & 4/15 \\
Hotel Niwa Tokyo & No & 0 & 1 & 0 & 1 & 1 & 3/15 \\
Marunouchi Hotel & No & 1 & 0 & 0 & 1 & 0 & 2/15 \\
Shibuya Stream Excel & No & 0 & 0 & 0 & 1 & 0 & 1/15 \\
\bottomrule
\end{tabular}
\end{table*}

Cited hotels average 8.6/15 (range: 5--13); control hotels average 3.4/15 (range: 1--5), a significant difference (Mann-Whitney $U = 48$, $p = 0.003$). Every hotel scoring 6 or above was cited; every hotel below 6 was not (Fisher exact $p = 0.002$).

The K5 case illustrates why depth matters more than feature presence. Hotel K5---a design boutique hotel in Nihonbashi with a 9.6/10 Booking.com rating, bilingual content, Schema.org markup, a neighborhood page, and a FAQ---scores only 5/15 on the depth scale because each feature is brief. Gemini mentions K5 by name in three responses but draws its information from OTA and editorial sources (Expedia, Hotels.com, myboutiquehotel.com), not from k5-tokyo.com. The hotel is discovered, but through intermediaries. By contrast, Kadoya Hotel---an independent with no schema markup and no technical SEO---scores 13/15 because its 33-question FAQ, 13-attraction sightseeing guide, and regular blog create 20+ indexed pages that directly answer traveler queries. Kadoya achieves direct citation; K5 does not.

The audit suggests a two-stage process for hotel direct citation in AI search: first, the hotel website must contain content deep enough to rank in Google Search for relevant queries; second, the retrieved content must answer the question better than competing OTA or editorial pages. Hotels that fail at the first stage never reach the second.

\section{Discussion}

\subsection{What This Means for Hotel Discovery}

Traditional hotel discovery rewards accumulated authority---review volume, ratings, advertising spend, domain authority. Our findings are consistent with a retrieval process in which question-content fit plays a major role, even if authority and ranking signals still matter. The intent-source divide (Section~3.1) supports this interpretation: when the question is transactional, non-OTA sources account for only 30.8\% of citations; when the question is experiential, non-OTA citation rises to 55.9\%. The 25.1 percentage-point swing suggests that the source of hotel discovery in AI search is not fixed---it depends on how the query is framed.

\subsection{The Hotel Website as Discovery Asset}

The most actionable finding for hotel operators is that hotel-direct citation is stable across query intent but sensitive to content depth. Three observations point in this direction, with the third remaining exploratory.

\textbf{First, hotel-direct citation does not vary by intent.} Hotel-direct rates are stable across transactional and experiential queries in both languages (Table~\ref{tab:hotel_direct}). This means a hotel website with deep, question-answering content earns citations under both conditions---it does not need to be optimized for one narrow class of prompts. The intent-stability is theoretically consistent with how grounding works: a hotel website with neighborhood guides, transit information, and atmosphere descriptions is relevant to both transactional and experiential queries simultaneously.
\smallskip\noindent\textbf{Second, Japanese queries produce significantly more hotel-direct citations than English queries} (Section~3.3: query-level OR = 2.27, $p = 0.030$). Japanese hotel brand websites tend to contain deeper question-answering content---neighborhood guides, transit information, area descriptions---than their English counterparts, which tend to be booking-oriented. The implication is that content depth at the language-ecosystem level translates into measurably higher hotel-direct citation rates.
\smallskip\noindent\textbf{Third, content depth---not just content presence---separates cited from non-cited hotels.} The search-answerable depth (SAD) audit of 14 hotels (Section~3.4) shows that cited hotels score significantly higher than non-cited hotels (mean 8.6/15 vs.\ 3.4/15; Mann-Whitney $U = 48$, $p = 0.003$). Every hotel scoring 6 or above on the depth scale was cited; every hotel below 6 was not (Fisher exact $p = 0.002$). The K5 case is instructive: a hotel with a perfect binary feature score (FAQ page, area guide, bilingual content) but shallow depth on each feature was never cited directly---Gemini drew K5 information from OTA and editorial intermediaries instead.

Hotels cannot control how travelers phrase queries, but they can control the depth of their own content. Our exploratory audit suggests that content depth may be one factor that distinguishes hotels whose websites capture the discovery moment from those discovered through intermediaries.

\subsection{Implications for Different Stakeholders}

\textbf{Independent hotels} may have the most to gain. The K5-vs-Kadoya comparison (Section~3.4) shows that brand prestige and technical sophistication do not determine direct citation---content depth does. Kadoya, a single independent with no schema markup and no SEO, achieves direct citation through a 33-question FAQ and a 13-attraction sightseeing guide. K5, a design boutique with a 9.6 rating and full technical implementation, does not. Content depth may level the playing field in ways that OTA rankings do not.

\textbf{Domestic chains} are well-positioned in Japanese but underinvested in English. Japanese queries produce 2.27$\times$ more hotel-direct citations than English queries (Section~3.3, $p = 0.030$), consistent with deeper Japanese-language content on domestic chain websites. For the 36.9 million inbound visitors to Japan \cite{jnto2025}, the English-language content gap represents unrealized discovery potential.

\textbf{International chains} have the resources for content investment but face structural barriers. Kimpton Shinjuku scores 13/15 on search-answerable depth and achieves direct citation---but Kimpton is an outlier. Most international chain property pages are centrally managed booking endpoints without the location-specific, question-answering content that AI search retrieves.

\textbf{OTAs} remain the dominant citation source at 55.3\% overall and 69.2\% for transactional queries. Disintermediation is not imminent. However, OTA citation drops to 44.1\% for experiential queries---a 25.1 percentage-point swing that represents the contestable frontier where non-OTA sources, including hotel-direct websites, can compete.

\subsection{Connection to the Broader GEO Field}

Our findings contribute to the emerging understanding of how AI search systems allocate visibility. The GEO literature has identified systematic patterns: citation concentration among a few outlets \cite{yang2025news}, earned media bias \cite{chenm2025geo}, geographic and commercial bias \cite{li2024generative}, and a discovery gap for new entities \cite{sharma2026discovery}. Our study adds a domain-specific dimension by showing that these patterns operate differently depending on query intent and the structure of the underlying web ecosystem.

The ``earned media bias'' identified by Chen, M. et al.~\cite{chenm2025geo}---a systematic preference for editorial content over brand-owned content---requires reinterpretation in light of our results. In the hotel domain, OTAs are commercial intermediaries, not earned media in the traditional sense. Our data suggest that AI search does not prefer earned media per se but rather cites whatever content type best answers the query: OTA listing pages for transactional queries, editorial and blog content for experiential queries. The apparent earned media bias may be partially an artifact of query distribution---if experiential queries predominate in a study, earned media will naturally appear favored. Our taxonomy reveals that the intent-source divide operates along the OTA/non-OTA boundary rather than the brand-owned/earned-media boundary. Hotel direct websites constitute a consistent 19--24\% of the non-OTA channel across intent conditions, while OTA sources fluctuate dramatically. This suggests that the relevant analytical frame for AI search citation is not ``earned vs.\ owned media'' but ``which content best answers the specific query.''

The role-augmented intent-driven GEO framework \cite{chenx2025role} provides theoretical scaffolding for this finding. That framework argues that query intent is a primary determinant of AI search behavior. Our four-category, two-intent design provides empirical evidence within the hospitality domain: the same hotel market produces dramatically different citation patterns depending on whether the query is transactional or experiential. The intent-source divide may reflect a structural property of how AI search matches queries to available content.

\section{Limitations}

This study has six limitations that bound the generalizability and causal interpretation of our findings.

\textbf{1. Single platform, single point in time.} All 156 queries were executed on Gemini 2.5 Flash with Google Search grounding in March 2026. Google regularly updates both the underlying language model and the grounding mechanism that determines which web sources are retrieved and cited. A model version change, a modification to grounding source ranking, or an update to Google Search's index could alter the citation patterns we document. Furthermore, other AI search platforms---ChatGPT with browsing enabled, Perplexity, Claude---employ different retrieval architectures and may exhibit different source-type preferences. ChatGPT uses Bing rather than Google Search for web grounding, which accesses a different index and applies different ranking signals. Perplexity maintains its own search index and citation system. Our findings are specific to the Gemini-Google Search grounding pipeline and should not be assumed to generalize across platforms without empirical cross-validation.

\textbf{2. Correlation, not causation.} We observe that hotels with richer website content---neighborhood guides, FAQ pages, sightseeing information, access maps---receive more direct citations from Gemini. However, we cannot establish that content depth causes direct citation. The observed association is susceptible to confounding: well-known hotels with strong brands may both invest in website content and receive citations due to brand recognition, search engine authority, or backlink profiles that are correlated with but distinct from content depth. A hotel that is frequently mentioned on third-party websites may accumulate domain authority that benefits its direct citation rate independently of its own content. The K5 case (Section~3.4) reinforces this limitation: a hotel with strong content features but shallow search-answerable depth was not cited, suggesting that content alone does not determine citation---Google Search ranking also plays a role. The causal test---modifying a hotel's website content and measuring subsequent changes in Gemini citation behavior---would require a controlled intervention design with pre-post measurement, which we propose as future work but have not conducted in this study.

\textbf{3. Single market.} Tokyo's hotel web ecosystem has characteristics that may limit generalizability to other tourism markets. The Japanese web contains a dense network of domestic OTAs (Jalan, Rakuten Travel, Ikyu), Japanese editorial platforms (icotto.jp, ozmall.co.jp), and Japanese-language hotel brand websites that collectively create a rich non-OTA content supply. Other tourism markets may have thinner non-OTA content ecosystems, which would affect the magnitude of the intent-source divide. The EN-JP difference in hotel direct citation rate (11.0\% vs.\ 8.2\%, $p = 0.090$) may be specific to the bilingual information landscape of Tokyo, where two linguistically separate web ecosystems serve the same geographic market. Cities with a single dominant language (e.g., Paris, Bangkok) or cities where the local-language web is less developed would present different dynamics. Additionally, Tokyo's hotel market includes distinctive property types (capsule hotels, ryokans, business hotels with specific Japanese amenities) that generate unique content not found in other markets, potentially inflating the experiential query effect.

\textbf{4. Query framing and answer-type specificity.} Our experiential queries are not only longer and lexically richer than transactional queries (addressed via robustness checks in Appendix~E) but also embed implicit constraints on what constitutes a good answer. A supplementary test with 20 ``experiential-but-OTA-answerable'' queries (Appendix~E) found that experiential framing produces high non-OTA rates (54.0\%) even when the question could be answered by OTA review data, suggesting the effect is driven by framing rather than answer-type bias. However, this supplementary set is small (20 queries, 137 citations) and was not part of the original study design. A larger, more systematic manipulation of answer-type specificity independent of framing would provide stronger evidence.

\textbf{5. Sample size and partial replication.} Our 156 queries yielded 1,357 citations, providing adequate statistical power for the core intent-source divide ($\chi^2 = 84.23$, $p = 4.40 \times 10^{-20}$) and the language interaction effect ($p = 0.015$). However, sub-analyses---particularly the category-by-language-by-intent breakdowns and the hotel-direct intent comparison within each language---operate on smaller cell sizes where statistical power is reduced. Gemini's grounding output is non-deterministic: re-running the same query produces partially different citation sets (within-query SD = 0.142 in our 20-query replication test; Appendix~E). While the ICC of 0.656 indicates moderate-to-good reliability and the intent-source divide replicated in all five runs, the replication covered only 20 of 156 queries. A full multi-run design re-running all 156 queries would provide tighter query-level estimates, particularly for sub-group analyses. Additionally, the 156 queries are generated from 16 template types (4 categories $\times$ 2 intents $\times$ 2 languages), with variation only in the ward name. Citations from queries sharing a template are not fully independent, which may understate standard errors in the logistic regression. Given the large effect size ($\chi^2 = 84.23$), template clustering is unlikely to alter the headline finding, but a mixed-effects model with template as a random effect would provide more conservative inference.

\textbf{6. Discovery, not booking.} This study measures who controls the discovery moment---which sources AI search cites---not who closes the booking. The billboard effect \cite{anderson2009billboard,ghose2012ranking} means OTA-cited hotels may still receive direct bookings. Our OTA citation rates should be interpreted as measures of discovery intermediation, not booking capture.

\section{Future Work}

Three extensions would most strengthen these findings. First, \textbf{cross-platform validation}---executing the same query set on ChatGPT, Perplexity, and Claude---would test whether the intent-source divide is specific to Gemini or a general property of AI search. Second, a \textbf{causal intervention study} partnering with a hospitality business to expand website content and track citation changes would establish whether content depth causes citation rather than merely correlating with it. Third, \textbf{multi-market replication} in cities with different OTA landscapes and web ecosystems would test generalizability beyond Tokyo.

\section{Conclusion}

AI search may be making hotel discovery more contestable. This study documents a 25.1 percentage-point shift in non-OTA citation between transactional and experiential queries, amplified in Japanese where a more diverse content ecosystem absorbs a larger share of the experiential shift. Hotel-direct citation is stable across intent conditions, and Japanese queries produce significantly more hotel-direct citations than English queries (OR = 2.27, $p = 0.030$), consistent with deeper question-answering content on Japanese hotel websites. For an industry that has historically rented discovery from commission-based intermediaries, these findings suggest that the discovery moment itself is becoming contestable. Hotels with deep, question-answering content appear better positioned to compete for that moment directly. Our results do not imply that AI will eliminate intermediation or that citation source maps onto booking channel. They do suggest that AI search opens a new pathway to discovery---one where hotels with deep, question-answering content can be found without paying an intermediary for the privilege.

\bibliographystyle{plain}
\bibliography{references}

\appendix
\onecolumn

\section{Experimental Setup}

\textbf{Model and API configuration.} All queries were executed on Gemini 2.5 Flash via the Google GenAI Python SDK (\texttt{google-genai}) with the following configuration:

\begin{itemize}
\item Model: \texttt{gemini-2.5-flash}
\item Google Search grounding: enabled (\texttt{GoogleSearch} tool)
\item Temperature: 1.0 (default; not explicitly overridden)
\item Seed: not available (Gemini API does not support deterministic seeding for grounded responses)
\item Data collection period: March 2026
\end{itemize}

\textbf{Response structure.} Each API response contains: (a) a generated text answer in markdown, and (b) a structured list of grounding citations (\texttt{grounding\_chunks})---the URLs and titles of web pages consulted during response generation. We extract and classify the grounding citations; the text responses are used only for hotel name extraction.

\textbf{Example query and response format:}

\begin{verbatim}
{
  "query_id": "q0157",
  "model": "gemini-2.5-flash",
  "prompt": "Cheap hotel in Chiyoda",
  "query_language": "en",
  "query_intent": "transactional",
  "query_category": "budget",
  "area": "Chiyoda",
  "sources": [
    {"url": "...", "text": "hotelscombined.com"},
    {"url": "...", "text": "expedia.com"},
    {"url": "...", "text": "hotels.com"},
    {"url": "...", "text": "hostelworld.com"},
    {"url": "...", "text": "kayak.com.ph"},
    {"url": "...", "text": "momondo.co.za"},
    {"url": "...", "text": "agoda.com"}
  ]
}
\end{verbatim}

\section{Sample Queries}

The following 10 queries illustrate the paired transactional-experiential design across categories and languages. The full corpus comprises 156 queries (144 area-level + 12 city-level).

\begin{table}[H]
\caption{Sample Queries from the Paired Design}
\label{tab:sample_queries}
\centering
\small
\begin{tabular}{lllll}
\toprule
ID & Language & Category & Intent & Query \\
\midrule
q0157 & EN & Budget & Transactional & Cheap hotel in Chiyoda \\
q0176 & EN & Budget & Experiential & Good value hotel with local charm in Chuo \\
q0194 & EN & Rating & Transactional & Best rated hotel in Chuo \\
q0212 & EN & Rating & Experiential & Hotel with exceptional service and atmosphere in Chuo \\
q0229 & EN & Convenience & Transactional & Hotel near Chiyoda station with easy access \\
q0169 & JP & Budget & Transactional & 新宿で安いホテル \\
q0184 & JP & Budget & Experiential & 東京都千代田区で地元の雰囲気が楽しめるコスパの良いホテル \\
q0208 & JP & Rating & Transactional & 豊島で評価の高いホテル \\
q0275 & JP & Business & Transactional & 東京都中央区で出張向けのホテル \\
q0293 & JP & Business & Experiential & 東京都中央区で仕事スペースがあり静かなホテル \\
\bottomrule
\end{tabular}
\end{table}

\section{Source Classification Taxonomy}

Citations were classified using domain-substring matching. The table below shows the taxonomy structure with representative examples.

\begin{table}[H]
\caption{Source Classification Taxonomy}
\label{tab:taxonomy}
\centering
\small
\begin{tabular}{lll}
\toprule
Category & Type & Examples \\
\midrule
\textbf{OTA} & Global OTA & booking.com, expedia.com (incl.\ regional variants), hotels.com, tripadvisor.com \\
 & Japanese OTA & jalan.net, rakuten.co.jp, ikyu.com, jtb.co.jp \\
 & APAC OTA & agoda.com, trip.com, traveloka.com \\
 & Meta-search & kayak.com, trivago, skyscanner, hotelscombined \\
\textbf{Non-OTA} & Hotel direct & kimptonshinjuku.com, superhotel.co.jp, princehotels.com \\
 & Editorial curation & wanderlog.com, myboutiquehotel.com, icotto.jp, ozmall.co.jp \\
 & Travel blog & Individual travel bloggers \\
 & Travel media & Forbes, Lonely Planet, livejapan.com \\
 & Local tourism & Ward-level tourism boards, gotokyo.org \\
 & Coworking/workspace & bizcomfort.jp, e-office.space \\
 & Travel agency & skyticket.jp, travel.co.jp \\
 & User-generated & YouTube, Reddit, note.com \\
 & Accommodation platform & Airbnb, HafH \\
\bottomrule
\end{tabular}
\end{table}

\section{Statistical Results}

\begin{table}[H]
\caption{Citation-Level Logistic Regression---$P(\text{non-OTA Citation})$}
\label{tab:logistic}
\centering
\small
\begin{tabular}{lcccccc}
\toprule
Variable & M1 Coef & M1 OR [95\% CI] & M1 $p$ & M2 Coef & M2 OR [95\% CI] & M2 $p$ \\
\midrule
Intercept & $-$1.429 & 0.24 [0.18, 0.32] & $<$ 0.001 & $-$1.241 & 0.29 [0.21, 0.40] & $<$ 0.001 \\
Intent (experiential) & 1.080 & 2.95 [2.34, 3.71] & $<$ 0.001 & 0.782 & 2.19 [1.57, 3.04] & $<$ 0.001 \\
Language (Japanese) & 0.283 & 1.33 [1.06, 1.67] & 0.015 & $-$0.053 & 0.95 [0.67, 1.35] & 0.769 \\
Category: Rating & 0.039 & 1.04 [0.74, 1.47] & 0.827 & 0.037 & 1.04 [0.73, 1.47] & 0.833 \\
Category: Convenience & 0.694 & 2.00 [1.48, 2.71] & $<$ 0.001 & 0.690 & 1.99 [1.47, 2.70] & $<$ 0.001 \\
Category: Business & 0.944 & 2.57 [1.87, 3.54] & $<$ 0.001 & 0.945 & 2.57 [1.87, 3.54] & $<$ 0.001 \\
Experiential $\times$ Japanese & --- & --- & --- & 0.571 & 1.77 [1.12, 2.80] & 0.015 \\
\bottomrule
\end{tabular}

\smallskip\noindent
Model~1: AIC = 1736.0, Log-L = $-$862.0. Model~2: AIC = 1732.1, Log-L = $-$859.1. LR test: $\chi^2(1) = 5.90$, $p = 0.015$. $n = 1{,}357$ citations. Reference: transactional, English, budget. DV = non-OTA citation.
\end{table}

\begin{table}[H]
\caption{Query-Level Quasibinomial Regression ($n = 156$ queries)}
\label{tab:quasibinomial}

\smallskip\noindent
To address non-independence, we re-estimated Model~2 at the query level using grouped binomial regression. Scale parameter = 0.262 (slight underdispersion).

\smallskip
\begin{tabular}{lcc}
\toprule
Variable & OR [95\% CI] & $p$ \\
\midrule
Experiential & 2.19 [1.85, 2.59] & $<$ 0.001 \\
Japanese & 0.95 [0.79, 1.14] & 0.566 \\
Rating & 1.04 [0.87, 1.24] & 0.680 \\
Convenience & 1.99 [1.71, 2.33] & $<$ 0.001 \\
Business & 2.57 [2.18, 3.03] & $<$ 0.001 \\
Experiential $\times$ Japanese & 1.77 [1.40, 2.24] & $<$ 0.001 \\
\bottomrule
\end{tabular}

\smallskip\noindent
Coefficients are substantively identical to citation-level estimates, confirming the finding is not an artifact of pseudo-replication.
\end{table}

\begin{table}[H]
\caption{Complete Statistical Test Summary}
\label{tab:test_summary}
\centering
\small
\begin{tabular}{llrl}
\toprule
Test & Statistic & Value & $p$ \\
\midrule
\multicolumn{4}{l}{\textbf{Core finding}} \\
Chi-squared (intent $\times$ source) & $\chi^2(1)$ & 84.23 & $4.4 \times 10^{-20}$ \\
Unadjusted OR (non-OTA, E vs T) & OR [95\% CI] & 2.84 [2.27, 3.56] & --- \\
Cramer's V & $V$ & 0.249 & --- \\
Mann-Whitney U (query-weighted) & $U$ & 4,721 & $< 10^{-9}$ \\
\midrule
\multicolumn{4}{l}{\textbf{Category-level}} \\
Budget & $\chi^2(1)$ & 26.00 & $3.4 \times 10^{-7}$ \\
Rating/Quality & $\chi^2(1)$ & 16.78 & $4.2 \times 10^{-5}$ \\
Business & $\chi^2(1)$ & 25.81 & $3.8 \times 10^{-7}$ \\
Convenience & $\chi^2(1)$ & 13.29 & $2.7 \times 10^{-4}$ \\
\midrule
\multicolumn{4}{l}{\textbf{Language interaction}} \\
LR test (Model~2 vs Model~1) & $\chi^2(1)$ & 5.90 & 0.015 \\
Interaction OR (experiential $\times$ JP) & OR [95\% CI] & 1.77 [1.12, 2.80] & 0.015 \\
Direct cell: JP-E vs EN-T (non-OTA, unadj.) & OR [95\% CI] & 3.52 [2.53, 4.90] & $< 10^{-14}$ \\
\midrule
\multicolumn{4}{l}{\textbf{Non-OTA composition}} \\
JP-domain share (JP queries) & \% & 68.4\% & --- \\
JP-domain share (EN queries) & \% & 6.4\% & --- \\
EN vs JP composition & $\chi^2(10)$ & 171.90 & $< 0.001$ \\
Hotel-direct EN vs JP & $\chi^2(1)$ & 2.87 & 0.090 (n.s.) \\
\midrule
\multicolumn{4}{l}{\textbf{Robustness}} \\
Answer-type specificity (non-OTA: OA vs E) & $z$ & $-$0.40 & 0.69 (n.s.) \\
Answer-type specificity (non-OTA: OA vs T) & $z$ & 5.14 & $< 0.0001$ \\
Query-length OLS (pooled, intent $\beta$) & $\beta$ & 0.27 & 0.0001 \\
Query-length OLS (JP only, intent $\beta$) & $\beta$ & 0.45 & $< 0.0001$ \\
ICC (test-retest, 20 queries $\times$ 5 runs) & ICC(1,1) & 0.656 & --- \\
\midrule
\multicolumn{4}{l}{\textbf{Hotel-direct}} \\
Hotel-direct EN vs JP (query-level logistic) & OR [95\% CI] & 2.27 [1.08, 4.75] & 0.030 \\
Hotel-direct JP-E vs EN-E (one-sided z) & $z$ & --- & 0.026 \\
\midrule
\multicolumn{4}{l}{\textbf{Exploratory}} \\
SAD cited vs control (Mann-Whitney) & $U$ & 48 & 0.003 \\
SAD threshold (Fisher exact) & --- & --- & 0.002 \\
\bottomrule
\end{tabular}
\end{table}

\section{Robustness Checks}

\textbf{Query length and lexical richness.} OLS regression controlling for query length, adjective density, subjectivity, and type-token ratio confirms that intent remains the strongest predictor in a pooled model ($\beta = 0.27$, $p = 0.0001$, $n = 156$). In Japanese alone, intent survives all lexical controls ($\beta = 0.45$, $p < 0.0001$).

\textbf{Answer-type specificity.} 20 supplementary queries using experiential language but asking OTA-answerable questions produced an OTA citation rate of 46.0\% (54.0\% non-OTA, $n = 137$), statistically indistinguishable from the original experiential non-OTA rate of 55.9\% ($z = -0.40$, $p = 0.69$) and significantly higher than the transactional non-OTA rate ($z = 5.14$, $p < 0.0001$).

\textbf{Test-retest reliability.} Re-running 20 stratified queries five times each yielded ICC(1,1) = 0.656 (moderate-to-good; \cite{cicchetti1994guidelines}). The intent-source divide was positive in all five runs.

\textbf{Query-weighted analysis.} Treating each query as a single observation: non-OTA rates 27.1\% (transactional) vs.\ 55.1\% (experiential). Mann-Whitney $U = 4{,}721$, $p < 10^{-9}$.

\end{CJK}
\end{document}